# Development of a high quality thin diamond membrane with embedded nitrogen-vacancy centers for hybrid spin-mechanical quantum systems

S. Ali Momenzadeh[†,*], Felipe Fávaro de Oliveira[†], Philipp Neumann[†], D. D. Bhaktavatsala Rao[†, §], Andrej Denisenko[†], Morteza Amjadi[¶], Sen Yang[†], Neil B. Manson[‡], Marcus W. Doherty[‡, **], and Jörg Wrachtrup[†, §]

[†] 3[rd] Institute of Physics, Research Center SCoPE and IQST, University of Stuttgart, 70569 Stuttgart, Germany

[§] Max Planck Institute for Solid State Research, Heisenbergstrasse 1, 70569 Stuttgart, Germany

[¶] Physical Intelligence Department, Max Planck Institute for Intelligent Systems, Heisenbergstrasse 3, 70569 Stuttgart, Germany

[‡] Laser Physics Centre, Research School of Physics and Engineering, Australian National University, Australian Capital Territory 0200, Australia

**ABSTRACT:** Hybrid quantum systems (HQSs) have attracted several research interests in the last years. In this Letter, we report on the design, fabrication, and characterization of a novel diamond architecture for HQSs that consists of a high quality thin circular diamond membrane with embedded near-surface nitrogen-vacancy centers (NVCs). To demonstrate this architecture, we employed the NVCs by means of their optical and spin interfaces as nanosensors of the motion of the membrane under static pressure and in-resonance vibration, as well as the residual stress of the membrane. Driving the membrane at its fundamental resonance mode, we observed coupling of this vibrational mode to the spin of the NVCs by Hahn echo signal. Our realization of this architecture will enable futuristic HQS-based applications in diamond piezometry and vibrometry, as well as spin-mechanical and mechanically mediated spin-spin coupling in quantum information science.

There have been several proposals[1-2] to exploit mechanical degrees of freedom to control and couple nitrogen-vacancy centers (NVCs) for applications in quantum information science[1-4] and nano-force sensing[5-6]. For example, mechanical motions of diamond microcantilevers have been detected via the electronic spins of NVCs[7-9]. Such proposals rely on the incorporation of NVCs into well-characterized nano/micro-mechanical structures that behave according to simple models from continuum mechanics. However, the realization of such ideal structures, free from complications like residual stress, is a significant challenge yet to be achieved. Thus, a vital first step is to comprehensively characterize these factors and their influence on the performance of the structures. Furthermore, unlike other materials such as SiN[10], the manufacturing of large, homogenous, and high-quality thin diamond membranes for the simple, robust and scalable fabrication of diamond nano/micro-mechanical structures is yet to be demonstrated.

In this Letter, we report a novel approach to the systematic design, fabrication, and characterization of a circular diamond membrane (hereinafter will be mentioned shortly as "membrane") incorporating NVCs at the average depth of $\approx$20 nm. The membrane has a diameter of $\approx$1.1 mm, a thickness of $\approx$1.2 µm and surface roughness of $\approx$0.4 nm. To examine the mechanical properties of the membrane, we employed the confocal microscopy technique that uses the fluorescence point spread function (PSF) of individual embedded NVCs as nanosensors to measure the deflection of the membrane under an applied pressure. In this way, and by applying the theory of thin membranes[11] we have inferred an effective thickness of $\approx$1.2 µm and average radial residual stress of 54 MPa. By means of the spin-mechanical coupling of the NVCs, we employed them as nanoprobes for the motion of the membrane under applied static pressure (DC) and in-resonance vibration (AC), as well as residual stress. In this way, we observed $\approx$2.3 MHz of spin resonance frequency shift under 1 bar of applied DC pressure, and coupling of the spin state to the vibrational state of the membrane detected by Hahn echo sequence showing negative offset. In addition, a radial residual stress of ≈52 MPa was measured by means of the spins of single NVCs, that shows good consistency with the value derived from the confocal microscopy measurements and model of the membrane's mechanics.

Our approach to fabricating the thin diamond membrane was as follows: NVCs were generated by aid of nitrogen implantation technique[12] in a ~2mm×2mm×0.027mm [100]-oriented CVD electronic grade diamond film. The implantation dose was chosen so resulting in single NVCs resolvable by confocal microscopy. Then, by means of a novel etching technique based on an $Ar/SF_6$ plasma gas mixture and an auxiliary diamond with an angled-wall hole to act as an etching mask, the membrane was created from the diamond film by etching for a depth of >25 μm (Figure 1b). A crucial role was played by the angled wall of the hole in the diamond etching mask (Figure 1a) which resulted in homogenous etching with minor deviations over the whole membrane area (~ 0.1%), as well as the avoidance of cracks and holes at the membrane edge, which appeared in earlier attempts. The presented novel etching recipe achieves both a relative high etching rate and a remarkable smooth surface. The etching rate was measured to be ≈170 nm/min followed by surface roughness of ≈0.4 nm showing ~5-fold enhancement in comparison to the initial value (measured by AFM[13]). More detailed information about the fabrication process is given in Supporting Information (SI).

As the first characterization step of the final structure, we performed optical measurements of the membrane deflection under a DC pressure, which in turn confirms the valid continuum mechanical model of the structure. To that aim, the sample was installed under a nitrogen gas pressure vessel (as depicted schematically in Figure 1c), positioned on top of a home-built confocal setup capable of pulsed optical and microwave pulses (as explained before[14]). The fluorescence PSF[15] (z- axis) of single NVCs near the center of the membrane was monitored as the nitrogen gas pressure was increased. The applied pressure vs. deflection of the membrane at its center is shown in Figure 2a. The obvious non-linear (cubic polynomial) behavior clearly shows that the final structure thoroughly obeys the mechanical theory of thin circular membranes[11]. Fitting the data using the model (refer to SI), yielded values of t=1.2±0.2 μm and $\sigma_0$=54±6 MPa for the effective thickness and the radial compressive residual stress[11], respectively. This derived thickness is in good agreement with the value of ≈1-2 μm, obtained by direct microscopy.

To assess the sensitivity of the presented hybrid device to pressure, we applied a small pressure while monitoring the fluorescence photon count rate of a single NVC at the center of the membrane.

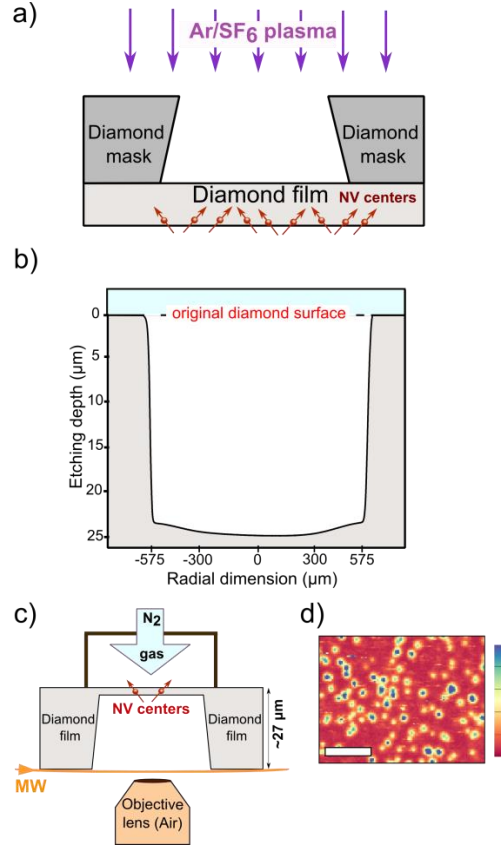

Figure 1. Fabrication and mounting scheme of the diamond thin membrane-NVC hybrid system. (a) Etching scheme using $Ar/SF_6$ plasma gas mixture and an auxiliary diamond etching mask is illustrated. (b) Depth profile of the diamond film including a thin membrane at the center; etching depth >25 μm was measured by Dektak instrument. The slightly thicker part at the rim is due to inhomogeneous etching of the film close to the diamond wall (SI). (c) Schematic of the sample and its mounting in the home-built confocal setup for DC measurements. (d) An exemplary confocal image of the NVCs near the center of the membrane. Scale bar shows 3 μm.

As can be seen in Figure 2b, a ≈2 × 10³ /sec decrease in the photon count rate was clearly observed. This change in the fluorescence rate can be then attributed to the displacement in z- focus of the PSF of the NVC and then converted into an applied DC pressure as small as ≈40 Pa. In this case, the photon shot noise limit on the pressure sensitivity is <6 $\frac{Pa}{\sqrt{Hz}}$, which can be improved by increasing the photon count rate of NVCs as nanoprobes through incorporating an ensemble of NVCs into this structure. Nevertheless, this geometry employing a single NVC can be still identified as a subtle broadband pressure sensing device[16], highlighting its potential for future HQS-based[17] piezometry applications.

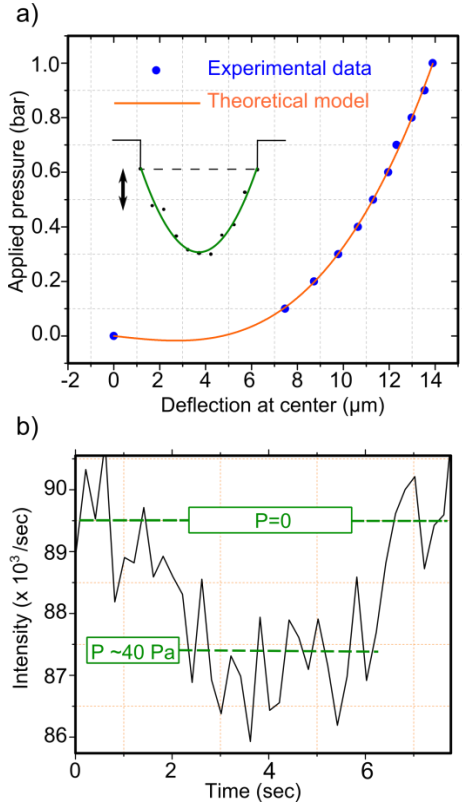

b)

Figure 2. Optical investigations of the membrane - NVC hybrid system are depicted. (a) Applied DC pressure to the system vs. deflection at its center is plotted. The cubic polynomial nature of the curve is a clear characteristic of the mechanics of a thin circular membrane. In inset, the schematic drawing of the membrane including measurement points and parabolic fit under static pressure of 0.8 bar in different radial coordinates is depicted. Vertical scale bar shows ~5 μm. (b) Recording the fluorescence rate of a single NVC at the center of the membrane in the absence and presence of a small applied pressure. As explained in the main text, this pressure is calculated to be ≈40 Pa.

In the next step, we employed the ground state spins of the NVCs as nanoprobes of the DC mechanical motion of the membrane. Deflection of the membrane induces crystal stress[11], which alters the spin resonance frequencies revealing longitudinal and transverse shifts[18-21]. This work is focused only on the longitudinal frequency shift measurements. To detect this frequency shift, optically-detected magnetic resonance (ODMR) measurements were performed resolving hyperfine splitting. To be confined to the longitudinal frequency shift and suppress the transverse one[7-8], approximately 90 G of magnetic field was aligned to the axis of the NVC under investigation. As plotted in Figure 3, a semi-linear behavior of the longitudinal frequency shift (plotted in absolute values) vs. the applied pressure is observed, approaching the longitudinal frequency shift of ≈2.3 MHz/bar.

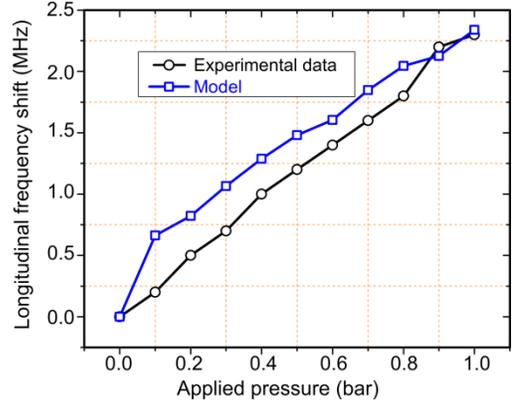

Figure 3. The longitudinal frequency shift (absolute value) is plotted vs. applied pressure. Our model, considering the membrane mechanics and spin-mechanical interactions, predicts a similar behavior of the system, in comparison to the values obtained by ODMR technique.

To provide a theoretical prediction for the ODMR longitudinal frequency shift vs. applied gas pressure, we applied the spin-stress model[6], which explains the longitudinal frequency shifts in terms of six different components of the stress tensor[22]. These components were calculated for different crystallographic defect orientations of NVCs in different positions of the membrane (a detailed calculation is provided in SI.) using our model of the membrane's mechanics and our optical deflection measurements. Considering our calculation, the maximum frequency shift occurs for NVCs at the center of the membrane. Assuming the thickness and radius of the membrane and Young's modulus[23] of diamond equal to 1.2 μm, 575 μm, and 1.2×10$^{12}$ Pa, respectively, the longitudinal frequency shift of a single NVC at the center of the membrane is calculated to be:

$$\Delta F_{||} \approx 0.012 \, (\text{MHz}/\mu m^2) \times w^2 \quad (1)$$

where w is the deflection at the center of membrane (in μm). As shown in Figure 3, an appropriate consistency between the experimental and theoretical curves can be seen, which supports our model of the spin-mechanical coupling and mechanics of the membrane. A slight deviation between these two curves can be attributed to the small inhomogeneity in the thickness of the membrane close to its rim as well as non-ideal clamping conditions.

In addition to the applied static pressure, single NVCs were employed also as local probes to sense the residual stress in the device. Residual stress is an inevitable component of micro- and nanostructures[11], which manipulates their mechanical properties up to a huge range[24]. As assumed in analytical continuum mechanics, the main residual stress in the membrane is the radial residual stress due to clamping[11]. Using ODMR technique, about fifteen NVCs per each four different crystallographic defect orientations near the center of the membrane were assessed and then compared by a control sample. Based on this measurement and using the spin-stress and continuum mechanics models, the average value of the residual stress was found to be 52 MPa, which is in a good agreement with the value of 54 MPa, obtained directly from optical measurements (more information is provided in SI.). This result highlights the capability of NVCs as sensitive nanoscale probes to characterize the residual stress in diamond mechanical structures.

Based on the presented investigations so far, the frequency shift measurement using ODMR technique shows a semi-linear behavior vs. applied pressure. However, the deflection measurement using optical technique delivers a non-linear output. Accordingly, it can be concluded that the optical interface is more sensitive in the range of low pressures (<0.5 bar), whereas the spin interface is likely to be more sensitive at higher pressures.

To study the AC mechanical behavior of the membrane, we focused on the lowest order (fundamental) vibrational mode that resembles a drum-shape vibration with maximum amplitude at the center of the membrane[11]. To excite the fundamental mode, the membrane was installed on top of a piezo chip connected to a frequency generator. The fluorescence photon count rate of a single NVC near the center of the membrane was then monitored while the frequency of the piezo chip was swept. A dip in the fluorescence intensity at $\omega \approx 2\pi \times 63$ kHz was detected, indicating the fundamental resonance mode of the membrane (refer to SI). The same result was observed using an interferometry method[24-25]. In both ways other technical errors, such as the resonance frequency of the piezo chip, were excluded by monitoring the thick part of the diamond lying far from the membrane. A 2-digit quality factor was obtained using both techniques, which is mainly due to atmospheric damping[27]. If required, this may be overcome by vacuum[25].

For resonant driving, the vibrational amplitude of the membrane was estimated optically via an analogous technique to the DC optical measurements. In this way, the vibrationally-induced broadening of the PSF of single NVCs near the center of the membrane was estimated vs. different piezo driving voltages. Like the displacement with DC pressure, the vibrational amplitude is non-linear in the piezo driving voltage. However, for small voltages (less than 2 V) it can be seen that the amplitude is approximately linear. In this voltage regime, a linear slope of ≈0.38 μm/V can be fitted (Figure 4a, more details can be found in SI).

To demonstrate the AC spin-mechanical coupling, the membrane's fundamental mode was again driven resonantly. A magnetic field of ≈290 G was aligned to the axis of NVCs at the center of membrane. NVCs near the center of the membrane couple dominantly to its fundamental vibration mode. This coupling results in a vibrationally-induced modulation of the spin resonances[3-4]. One can approximate this vibrationally-induced interaction between the NVC spin and the membrane oscillator in the linear regime (low voltages) through a periodic detuning term $\lambda \widetilde{z(v)} \cos\omega t$, where $\lambda$ is the coupling parameter of the spin to a membrane vibration quantum, and $\widetilde{z(v)}$ is the voltage-dependent dimensionless amplitude of the oscillation (i.e. vibration amplitude at each piezo driving voltage divided by the zero-point fluctuation of the membrane[4]). As shown by Bennett et al.[4], this non-synchronized vibrationally-induced detuning results in decoherence of the NVC spin, observable e.g. in Hahn echo signal. In this way, the Hahn echo signal depends upon the zeroth-order Bessel function[4]:

$$J_0\left[4\frac{\lambda \widetilde{z(v)}}{\omega}\sin^2\left(\frac{\omega t}{2}\right)\right] \qquad (2)$$

We observed experimentally (Figure 4b) that under driving of the membrane at its fundamental vibrational mode, the standard Hahn echo signal features a negative offset and shows an oscillation which fits to the Bessel function mentioned above (eq. 2). Using this fit, the coupling parameter can be estimated to be $\lambda = 6.12 \pm 0.51 \times 10^{-4}$ rad/s, which can be drastically improved by reducing the membrane radius, if desired[7]. More detailed information is provided in SI. The resulting driven decoherence rate ($\Gamma_{driven} = 2\pi/T_2^{driven}$) vs. piezo driving voltage is depicted in Figure 4c (more details can be found in SI).

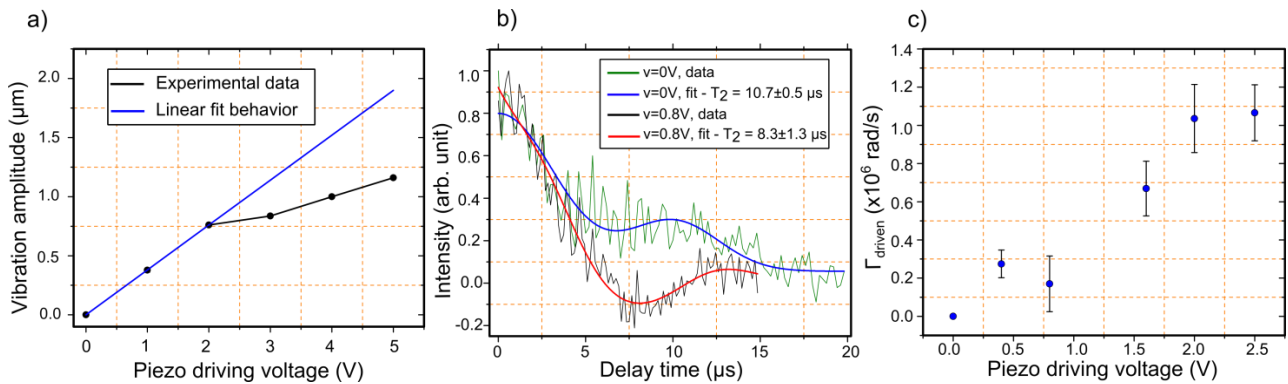

Figure 4. Optical and Hahn echo measurements in the membrane AC regime. (a) In-resonance vibration amplitude of the membrane vs. the piezo driving voltage was estimated by means of single NVCs, revealing an overall nonlinear behavior. Nevertheless, this behavior can be approximated to a linear fit with slope of 0.38 μm/V for voltages <2 Volts. (b) This figure demonstrates the Hahn echo signal under ≈290 G in absence (green curve with blue fit[29]) and presence (black curve with red fit) of in-resonance mechanical driving. A negative offset in the black curve following fluctuation behavior obeying Bessel function proves that our system is in driven regime, as presented theoretically before. (c) Driven decoherence rate is plotted vs. the driving voltage of the piezo chip. Nonlinear behavior of the membrane oscillator and possible heating effects can cause ascending decoherence rate.

As shown, it presents an ascending behavior as piezo driving voltage increases. The precise origin of this voltage-dependent decoherence is not clear. However, we suspect it to be related to the non-linearity of the membrane's mechanics and related factors, such as imperfect damping and coupled excitation of multiple vibrational modes. Further investigations would be beyond the scope of this work.

In conclusion, we demonstrated a versatile novel hybrid system based on a diamond thin circular membrane and embedded single NVCs. A novel etching technique was key to producing this hybrid device. Within this architecture, NVCs were employed as nanoscale probes to detect the membrane motion in DC and AC regimes, as well as residual stress of the structure. Theoretical calculations were also provided which supported the achieved experimental observations via NVC nanoprobes. It was seen that the optical interface of NVCs is likely more sensitive in the range of low applied DC pressures (<0.5 bar), whereas the ODMR technique is likely to be more sensitive at higher pressures. The presented results in general prove excellent capability of NVCs to be served in futuristic HQS-based piezometry and vibrometry applications.

Moreover, such a hybrid system can be exploited in optomechanical cavities utilizing NVCs, for instance in membrane-in-the-middle tries[10,26]. Furthermore, the presented system can be used in fluid mechanics applications as shown recently by diamond nanocantilevers[28]. This work was confined within the ground state spin levels associated with single NVCs, however excited state spin levels offer further opportunities[30]. One can benefit from the controllable stress in such a system to manipulate the spin levels in the excited state of NVCs[31], which is the backbone of their application in quantum information processing at low temperature[32-33].

## SUPPORTING INFORMATION

Detailed description of the sample fabrication and different part of experiments as well as theoretical studies is provided. This material is available free of charge via the Internet at http://pubs.acs.org.

## AUTHOR INFORMATION

### Corresponding Authors

* E-mail address:  a.momenzadeh@physik.uni-stuttgart.de
** E-mail address: marcus.doherty@anu.edu.au

### Notes

The authors declare no competing financial interest.

## ACKNOWLEDGMENT

We acknowledge Thomas Wolf, Thomas Reindl, Roland Nagy, Monika Ubl, Roman Kolesov, Nikolas Abt, Torsten Rendler, and Matthias Widmann for fruitful discussions. This work was supported by the EU via DIADEMS and SQUTEC, the DFG via FOR 1493 and SPP 1601, and the ARC (DP140103862). F.F. appreciates the financial support by CNPq project number 204246/2013-0.

## ABBREVIATIONS

HQS, hybrid quantum system; NVC, nitrogen-vacancy center; ODMR, optically-detected magnetic resonance;